\documentclass[twocolumn,showpacs,amsmath,amssymb, superscriptaddress]{revtex4}

\usepackage{amsmath}
\usepackage{amssymb}
\usepackage{amsfonts}

\usepackage{graphicx}
\usepackage{float}
\usepackage{color}

\usepackage{amsthm}

\DeclareMathOperator{\Tr}{Tr}

\newcommand{\angl}[1]{{\left\langle #1 \right\rangle}}

\makeatletter
\def\loweq@align#1#2{\lower.6ex\vbox{\baselineskip\z@skip\lineskip\z@
    \ialign{$\m@th#1\hfil##\hfil$\crcr#2\crcr=\crcr}}}
\def\lowsim@align#1#2{\lower.6ex\vbox{\baselineskip\z@skip\lineskip\z@
    \ialign{$\m@th#1\hfil##\hfil$\crcr#2\crcr\sim\crcr}}}
\def\geqq{\mathrel{\mathpalette\loweq@align >}}
\def\leqq{\mathrel{\mathpalette\loweq@align <}}
\def\grsim{\mathrel{\mathpalette\lowsim@align >}}
\def\lesssim{\mathrel{\mathpalette\lowsim@align <}}
\def\gsim{\mathrel{\mathpalette\lowsim@align >}}
\def\lsim{\mathrel{\mathpalette\lowsim@align <}}
\makeatother
\newcommand{\grless} 
{ {\, \raise-.24em\hbox{$<$} \hspace{-0.8em} \raise.31em\hbox{$>$}\, } }
\newcommand{\lessgr} 
{ {\, \raise-.24em\hbox{$>$} \hspace{-0.8em} \raise.31em\hbox{$<$}\, } }
\newfont{\bg}{cmr10 scaled\magstep4}                    
\newcommand{\bigzerou}{\smash{\lower1.7ex\hbox{\bg 0}}}

\newcommand{\crl}[1]{[-\infty,\infty]}

\newcommand{\ket}[1]{|{#1}\rangle}

\newcommand{\bra}[1]{\langle{#1}|}

\newcommand{\Ref}[1]{(\ref{#1})}

\newcommand{\da}[1]{#1^\dag}

\newcommand{\av}[1]{\langle#1\rangle}

\begin{document}

\title{Brachistochrone of Entanglement for Spin Chains}
  \author{Alberto Carlini}
 \email{alberto.carlini@uniupo.it}
 \affiliation{Dipartimento di Scienze ed Innovazione Tecnologica, Universita' del Piemonte Orientale, Alessandria, Italy}
 \affiliation{Istituto Nazionale di Fisica Nucleare, Sezione di Torino, Gruppo Collegato di Alessandria, Italy}
 \affiliation{NEST, Istituto di Nanoscienze-CNR, Pisa, Italy}

\author{Tatsuhiko Koike}
 \email{koike@phys.keio.ac.jp}
 \affiliation{Department of Physics and REC for NS, Keio University,
 Yokohama, 223-8522, Japan}

\date{December 1, 2016}
\begin{abstract}
We analytically investigate the role of entanglement 
in time-optimal state evolution 
as an application of the quantum brachistochrone, a general method for
obtaining the optimal time-dependent Hamiltonian for reaching a target quantum state. 
As a model, we treat two qubits indirectly 
coupled through an intermediate qubit 
that is directly controllable, 
which represents a typical situation in quantum information processing. 
We find the time-optimal
unitary evolution law and quantify residual entanglement by the 
two-tangle between the indirectly coupled qubits, 
for all possible sets of initial pure quantum states of a tripartite system.
The integrals of the motion of the brachistochrone are determined by fixing the minimal time at which the residual entanglement is maximized.   
Entanglement plays a role for $W$ and $GHZ$ initial quantum states, and for the bi-separable initial state in which the indirectly coupled qubits have a nonzero value of the 2-tangle.
\end{abstract}  

\pacs{03.67.-a, 03.67.Lx, 03.65.Ca, 02.30.Xx, 02.30.Yy}

\maketitle

\section{Introduction}

The concept of entanglement is one of the key features distinguishing the quantum world from the classical world, as
it captures those correlations which cannot have a classical origin \cite{epr}.
It is one of the fundamental resources in quantum information and computation theory (e.g., teleportation \cite{bennett1}, superdense coding \cite{bennett2},
quantum cryptography \cite{ekert}, quantum state tomography \cite{beck}, quantum repeaters \cite{sangouard}, quantum metrology \cite{giovannetti1}, quantum phase transitions \cite{amico}, black hole physics \cite{borsten} etcÉ).
Several measures of entanglement have been proposed for multipartite quantum systems in different contexts, for a review see, e.g., \cite{chuangnielsen}-\cite{bengtsson}.
On the other hand, the speed of evolution for quantum systems is an important concept, not only for determining the
theoretical limits at which quantum information can travel \cite{margolus}-\cite{levitin}, but also for the practical task of building quantum computers capable of performing fast quantum algorithms before the ubiquitous and disruptive decoherence effects come into play.
The importance of the connection between quantum entanglement, the speed of evolution of quantum systems and dynamical optimization problems has been discussed in \cite{giovannetti2}-\cite{kupferman}.
Furthermore, quantum optimal control is also a fundamental subject, both theoretically and experimentally, in quantum computing and information (see, e.g., \cite{brif} and the road map traced in the recent review \cite{glaser3}).
In particular, time-optimal quantum computation,  where the cost to be optimized is the time to achieve a given quantum evolution is relevant for the design of fast elementary gates and provides a more physical ground to describe the complexity of quantum algorithms.
A theoretical framework for the quantum brachistochrone (QB) was introduced in \cite{pure}.
The QB \cite{brach} is based on a variational principle enforcing the time-optimal evolution of a quantum system whose Hamiltonian is subject to a set of constraints (e.g., a finite energy, certain qubit interactions are forbidden etc..) and defines a boundary value problem with fixed initial and final quantum states (or unitary transformations). 
The QB has been studied for quantum state evolution in the case of pure \cite{pure} and mixed states
\cite{mixed}, for the optimal realization of unitary transformations up to a given target quantum gate \cite{unitary},
for the more realistic situation where the target can be reached
within a finite, tolerable error (a fixed fidelity) \cite{carlini3}, and for the time optimal transfer of coherence in a trilinear Ising spin chain \cite{carlini2}.
An efficient numerical algorithm for the QB was proposed and applied for 
finding evidence that the QB can be used to estimate 
the gate complexity of unitary operators \cite{ko}. 
It was also shown \cite{wang} that the QB can be recast as the problem of finding geodesics in a suitable Hilbert space and can be used to solve the famous Zermelo navigation
problem via the aid of a Finslerian geometry \cite{russell}.
Very recently \cite{geng} experimental time optimal universal control of spin qubits from diamonds with NV centers has been finally demonstrated based on the QB formalism.
On the other hand, the engineering of the generation and the dynamics, including the effects of (sudden) death and revival of tripartite entanglement for the prototypical $W$ 
or $GHZ$ states has been widely studied, e.g. with QED in optical cavities under the action of Markovian or non Markovian noise in 
\cite{bina}
-\cite{behzadi}, or within generic three qubit systems, e.g. 
\cite{buscemi}
-\cite{kalaga}.
Fast and robust engineering of tripartite entanglement from the vibrational modes in optomechanical systems can be found in \cite{xu}, while the generation of $W$ and $GHZ$ states has been studied, e.g., with the technique of shortcut to adiabatic passage, both using dressed states with SQUID qubits \cite{kang} or atoms in cavity QED \cite{chen}.
The dynamics of tripartite entanglement has been recently studied even in the context of gravity for accelerated qubits coupled to a real scalar field in the vicinity of black holes \cite{dai1}-\cite{dai2}, and for multiple identical fermions undergoing decoherence \cite{valdes}.
Quasi local control protocols for a dissipative engineering of $GHZ$ and $W$ states can be found in, e.g., \cite{ticozzi}, while the generation and stabilisation of the same states
in superconducting circuit QED via quantum feedback control techniques is shown in \cite{huang}.
Finally, optimal control methods to generate entanglement have been recently applied to continuous variable systems, e.g., for two mode systems in linear networks \cite{jacobs} or for two mode Gaussian states subject to Ohmic relaxation \cite{stockburger}, for plasmonically coupled quantum dots in cavity QED \cite{otten} and for noise resistant spin-squeezing in strongly interacting many body systems \cite{pichler}.
Recently, some authors \cite{borras2}-\cite{borras3} started the study of the role of quantum entanglement during the QB 
evolution of multipartite distinguishable systems in a pure quantum state, finding that the entanglement is 
pivotal to the QB evolution if at least two subsystems actively evolve.
Efficient generation of random multipartite entangled states has been also analyzed with the aid of time-optimal unitary operations \cite{borras4}.
More recently, the authors of \cite{zhao} found that genuine tripartite entanglement is necessary during the QB evolution of 
a set of three qubits in the pure state, except for the case in which less than three qubits attend evolution.

In this paper, we discuss the role played by entanglement in 
time-optimal state evolution, as an application of 
the QB formalism. 
Though our method can be applied to general systems, 
we consider a simple concrete model where two qubits are coupled via 
an intermediate qubit which can be controlled directly. 
This is a typical situation 
in a wide array of promising (scalable) experimental realizations of
quantum information processing (see, e.g.,
\cite{khaneja}-\cite{vandamme}). 
For example, the system can be considered as a spin chain under on-site
magnetic fields. 
An advantage of the model is that one can analytically work out all the details 
including the time evolution of the state and it entanglement. 
Moreover, the QB formalism naturally allows for the situation in which
local coherent controls are assumed to be time consuming, contrary to the
requirements of zero time cost for local controls typical of the
standard time-optimal quantum control methods \cite{khaneja1}-\cite{khaneja2}.
We concentrate our attention on the behavior of the residual entanglement between the indirectly coupled qubits at the 
end of a trilinear chain, as expressed by the so called 2-tangle \cite{coffman}.
We also consider all the possible initial quantum pure states for the tripartite system, i.e. completely separable, bi-separable, and with true tripartite entanglement (fully bipartite for $W$ and maximally tripartite for $GHZ$ states).
We then let the 2-tangle evolve along the general time-optimal quantum trajectory defined by the QB action principle, 
and  we fix the integrals of the motion by imposing that entanglement reaches its maximum value in the shortest possible time. The minimal time for reaching such a maximum of the 2-tangle is a function of the ratio between the interaction
couplings of the Ising Hamiltonian.

The paper is organized as follows.
In Section II we review the main features of the QB formalism for the time-optimal synthesis 
of unitary quantum evolutions.
In Section III we summarize the QB solution for the problem of a three-linear qubit system subject to an Ising interaction 
with unequal couplings and a local control on the intermediate qubit, when a finite energy is available.
In Section IV we define the measure for the bipartite entanglement between the two indirectly coupled qubits of the chain
and we introduce the main formulas for the  2-tangle and 3-tangle for all the possible sets of initial quantum states.
Section V is devoted to the study of the entanglement evolution for these initial quantum states, and we define the optimal
times and the analytical form of the quantum evolutions for which the 2-tangle between the indirectly coupled qubits 
most rapidly reach its maximum value.
Finally, Section IV is devoted to the summary and discussion of our results. 

\section{Time-optimal unitary evolution}

The goal is to determine the time-optimal way to generate the unitary evolution up to a certain $U_f$ (modulo physically irrelevant overall phases) by controlling an Hamiltonian $H(t)$ obeying the Schr\"odinger equation. 
We assume that $H$ is controllable and that only a finite energy is available in the experiment. 
This time-optimality problem may be formulated using the action \cite{unitary}:
\begin{align}
  \label{eq-action}
  S(U,H; \alpha, \Lambda,\lambda_j) &:=\int_0^1 ds \left[N \alpha +L_S+L_C\right],
  \\
  \label{eq-LS}
  L_S &:=\av{\Lambda,i \tfrac{dU}{ds}\da U- \alpha H},
  \\
  \label{eq-LC}
  L_C &:= \alpha   \sum_j{\lambda_j}f^j(H), 
\end{align}
where $\angl{A,B}:=\Tr (\da AB)$ and the Hermitian operator
$\Lambda(s)$ and the real functions $\lambda_j(s)$ are Lagrange multipliers.
The quantity $\alpha$ is the time cost, relating the parameter time $s$ and the physical time $t$ via $t:=\int \alpha(s) ds$.
Variation of $L_S$ by $\Lambda$ gives the Schr\"odinger equation:
\begin{align}
  \label{eq-Sch}
  i\frac{dU}{dt}=HU, \quad\text{or}\quad 
  U(t)={\mathcal  T}e^{-i\int^t_0 Hdt}, 
\end{align}
where $\mathcal T$ is the time ordered product. 
Variation of $L_C$ by $\lambda_j$ leads to the constraints for $H$:
\begin{align}
  f_j(H)=0. 
\label{constraints}  
\end{align}
In particular, the finite energy condition for a system of $\log N$ qubits reads: 
\begin{align}
f_0(H) :=\tfrac{1}{2}[\Tr(H^2)- N\omega^2]=0,
\label{eq-normH}
\end{align} 
where $\omega$ is a constant.

From the variation of $S$ with respect to $H$ we get:
\begin{align}
\Lambda = \lambda_0 H + \sum_{j\not = 0} \lambda_j \frac{\partial f_j(H)}{\partial H},
\label{eq-H}
\end{align} 
while from the variation of $S$ by $\alpha$ we obtain the normalization
condition:
\begin{align}
\Tr (H\Lambda)=N.
\label{eq-alpha}
\end{align}

Finally, variation of $S$ by $U$, use of eq. \Ref{eq-H} and some elementary algebra give the 
{\em quantum brachistochrone equation}: 
\begin{align}
  \label{eq-fund}
  i\frac{d\Lambda}{dt}= [H, \Lambda],
\end{align}
The {\it quantum brachistochrone} \Ref{eq-fund} together with the constraints \Ref{constraints} define a boundary-value problem for the evolution of the unitary operator $U(t)$ with fixed initial ($U(t=0)=1$, where $1$ is the identity matrix) and final conditions $(U(t=T)=U_f$, where $T$ is the optimal time duration necessary to achieve the target gate $U_f$).
It can be solved together with the constraint functions $f_j(H)$ to obtain $H_{\mathrm{opt}}(t)$.
Then one integrates the Schr\"odinger equation \Ref{eq-Sch} with $U(0)=1$ to get $U_{\mathrm{opt}}(t)$ and 
finally the integration constants in $H_{\mathrm{opt}}(t)$ can be fixed, e.g., by imposing that $U_{\mathrm{opt}}(T)$ 
equals a target $U_f$ modulo a global (physically irrelevant) phase.

\section{QB for a trilinear Ising chain} 

We now apply the general formalism of the QB to the case of a physical system of three qubits 
(labeled by  a
superscript $a\in \{1, 2,  3\}$) interacting via an Ising Hamiltonian and where the intermediate qubit is subject  to a local and controllable magnetic field $B_i(t)$ ($i=x, y, z$): 
\begin{align}
 H(t) :=J_{12}\sigma_z^{1}\sigma_z^{2} +J_{23}\sigma_z^{2}\sigma_z^{3} +\vec{B}(t)\cdot \vec{\sigma}^{2},
\label{}
\end{align}
where we have defined $\sigma_i^1\sigma_j^2:=\sigma_i\otimes \sigma_j\otimes 1$,
$\sigma_i^2\sigma_j^3:=1\otimes \sigma_i\otimes \sigma_j$,
$\sigma_i^{2} :=1\otimes \sigma_i\otimes 1$ and $\sigma_i$ are the Pauli
operators.
Introducing the ratio between the Ising interaction couplings $K:=J_{23}/J_{12}$ and rescaling time as $\tau:=J_{12}t$, the energy as $\hat \omega:=\omega/J_{12}$ and the magnetic field as  $\vec{B}_{\mathrm{opt}}(t)=J_{12}\vec{\hat B}_{\mathrm{opt}}(\tau)$, it can be shown \cite{carlini3} 
that the QB is solved by the following time-optimal magnetic field:
\begin{align}
\vec{\hat B}_{\mathrm{opt}}(\tau) = \left ( \begin{array}{c}
 {\hat B_0}\cos \theta(\tau)\\
 {\hat B_0}\sin \theta(\tau)\\
  \hat B_z   
  \end{array}\right ),
  \label{bopt}
\end{align}
where $\theta(\tau):=\hat\Omega \tau +\theta_0$ and $\hat\Omega$, $\theta_0$, $\hat B_0$ and $\hat B_z$ 
are integration constants.
The magnetic field is precessing around the $z$-axis with the frequency $\hat \Omega$.
Furthermore, the energy constraint explicitly reads $\vec{\hat B}^2=\hat \omega^2 -(1+K^2):=\hat\omega^2_K$, 
so that we can reparameterize:
\begin{align}
\hat B_0:=\hat\omega_K\cos\phi ; ~~~
\hat B_z:= \hat\omega_K \sin \phi,
\end{align}
where  and $\phi\in [0, 2\pi]$.

One can then solve the Schr\"odinger equation \Ref{eq-Sch} and find the time-optimal evolution operator (for more details, see Appendix B of \cite{carlini3}):
\begin{align}
U_{\mathrm{opt}}(\tau)&=e^{-i\frac{\hat \Omega \tau}{2}} A_D^{13 -}(\tau) {\ket{0}}_2\bra{0} + e^{ i\frac{\hat \Omega \tau}{2}} A_D^{13 +} (\tau) {\ket{1}}_2\bra{1}
\nonumber \\
& -i B_0 S_D^{13} (\tau) \biggl [ e^{i \left (\frac{\hat \Omega\tau}{2}-\theta(\tau)\right  )} {\ket{0}}_2\bra{1} + \mathrm{H.c.}\biggr ],
\label{uopt}
\end{align}
where we have introduced the diagonal operators acting in the Hilbert space of qubits 1 and 3:
\begin{align}
A_D^{13 \pm}(\tau)&:=\mathrm{Diag}[\alpha^{\pm}_{1}(\tau), \alpha^{\pm}_{2}(\tau), \alpha^{\pm}_{3}(\tau), \alpha^{\pm}_{4}(\tau)],
\\
S_D^{13}(\tau)&:=\mathrm{Diag}[s_{1}(\tau), s_{2}(\tau), s_{3}(\tau), s_{4}(\tau)].
\label{as}
\end{align}
The latter operators depend upon the functions of time: 
\begin{align}
s_{i}(\tau)&:=\frac{\sin (\omega_{i} \tau)}{\omega_{i}},
\label{si}
\\
c_{i}(\tau)&:=\cos (\omega_{i} \tau),
\label{ci}
\\
\alpha^{\pm}_i(\tau)&:=c_i(\tau) \pm i~b_is_i(\tau),
\label{ai}
\end{align}
and on the constants:
\begin{align}
\omega_i&:=\hat\omega_K\sqrt{\cos^2\phi +b^2_{i}},
\nonumber \\
b_i&:=\sin\phi +\frac{1}{\hat\omega_K}\biggl [(\delta_{i1}-\delta_{i4})(1+K)
\nonumber 
\\
&+(\delta_{i2}-\delta_{i3})(1-K)- \frac{\hat\Omega}{2}\biggr ],
\label{bi}
\end{align}
where $i=1,2,3,4$ and $\delta_{ij}$ is the Kronecker symbol.

\section{Entanglement in the 1-3 subsystem}

In this section, we start 
the analysis of the role played by the entanglement in time-optimal quantum
state evolutions. 
We consider the time-optimal evolution of an arbitrary initial pure state $\ket{\psi(0)}$ 
driven by the unitary operator \Ref{uopt} 
and study the behavior of the bipartite entanglement between indirectly coupled qubits 1 and 3.
We are interested, in particular, in determining 
the optimal time $\tau_*$ and the integration constants 
$\hat \Omega, \theta_0, \phi$ 
for which the 
the entanglement between qubits 1 and 3 
is maximized.

A pure state 
of a tripartite quantum system, 
with each party being a qubit, 
can be written as: 
\begin{align}
  \ket{\psi}=\sum_{i=0}^7a_i\ket{i}, 
\end{align}
where $\ket{0}:=\ket{000}$, $\ket{1}:=\ket{001}$, $\ket{2}:=\ket{010}$,
$\ket{3}:=\ket{011}$, $\ket{4}:=\ket{100}$, $\ket{5}:=\ket{101}$,
$\ket{6}:=\ket{110}$, $\ket{7}:=\ket{111}$. 
The entanglement between two of its subsystems, e.g. 1 and 3, may be 
quantified by the 2-tangle \cite{borsten}, \cite{coffman}:
\begin{align}
\tau_{13}:=2(\mathrm{Det}[\rho_1]-\mathrm{Det}[\rho_2]+\mathrm{Det}[\rho_3]
-|\mathrm{HypDet}(a)|),
\label{13tangle}
\end{align}
where $\rho_i$ ($i=1, 2, 3$) are the reduced density matrices: 
\begin{align}
\rho_1:=\Tr_{23}(\ket{\psi}\bra{\psi)}),
\nonumber \\
\rho_2:=\Tr_{13}(\ket{\psi}\bra{\psi)}), 
\nonumber \\
\rho_3:=\Tr_{12}(\ket{\psi}\bra{\psi)}), 
\label{rho123}
\end{align}
and $\mathrm{HypDet}(a)$ is Cailey's hyperdeterminant for the matrix of the coefficients $a_i$s (see eq. \Ref{hypdet} 
in the Appendix).

One may also define the 3-tangle:
\begin{align}
\tau_{123}:=|\mathrm{HypDet}(a)|,
\label{123tangle}
\end{align}
which describes the amount of  tripartite entanglement between all the spins.
 
The 
pure states of a tripartite quantum system can be classified into equivalence classes under local operations and 
classical communication (LOCC), 
which are distinguished by 
the degree of entanglement between its subsystems \cite{dur}. 
 In particular, we will consider the representatives of each of these classes as a possible initial quantum state, i.e.:

S) completely separable states, for which all 2-tangles and the 3-tangle vanish:
\begin{align}
\ket{\psi (0)}=\ket{\psi_S}:=\ket{000};
\label{sep}
\end{align}

B) bi-separable states, with bipartite entanglement, for which one of the 2-tangles is nonzero while the other 2-tangles and the 3-tangle are zero:
\begin{align}
\ket{\psi(0)}=\ket{\psi_{B1}}:=\frac{1}{\sqrt{2}}(\ket{001}+\ket{010}),
\\
\ket{\psi(0)}=\ket{\psi_{B2}}:=\frac{1}{\sqrt{2}}(\ket{001}+\ket{100}),
\label{eq-B2}
\\
\ket{\psi(0)}=\ket{\psi_{B3}}:=\frac{1}{\sqrt{2}}(\ket{010}+\ket{100});
\label{bisep}
\end{align}

W) W states, with full bipartite entanglement, for which all 2-tangles are non zero while the 3-tangle vanishes:
\begin{align}
\ket{\psi(0)}=\ket{\psi_{W}}:=\frac{1}{\sqrt{3}}(\ket{001}+\ket{010}+\ket{100});
\label{w}
\end{align}

GHZ) GHZ states with maximal tripartite entanglement, for which all 2-tangles and the 3-tangle are nonzero:
\begin{align}
\ket{\psi(0)}=\ket{\psi_{GHZ}}:=\frac{1}{\sqrt{2}}(\ket{000}+\ket{111}).
\label{ghz}
\end{align}

The time-optimal evolution operator $U_{\mathrm{opt}}(\tau)$, eq. \Ref{uopt}, drives the initial states according to:
\begin{align}
\ket{\psi(\tau)}=U_{\mathrm{opt}}(\tau)\ket{\psi(0)}=\sum_{i=0}^7a_i(\tau)\ket{i}.
\label{psitau}
\end{align}
For each of the possible classes of initial states, 
one can calculate the 
tangles 
$\tau_{13}(\tau)$ and $\tau_{123}(\tau)$ exactly. 
Namely, one computes the time-dependent amplitudes $a_i(\tau)$ through 
\eqref{psitau}, 
whose nonvanishing values are 
given by formulas \Ref{asep}-\Ref{aghz} in the Appendix, 
and 
substitutes these formulas into \Ref{hypdet} and \Ref{detrho123}, and then into \Ref{13tangle}
and \Ref{123tangle}. 

We obtained the following results. 

S-B1-B3) 
When 
there is no entanglement between the indirectly coupled qubits 1 and 
3 initially, i.e. when  the state $\ket{\psi(0)}$ is fully separable or biseparable of type
$B1$ or $B3$, 
both the $\tau_{13}(\tau)$ and the $\tau_{123}(\tau)$ tangles are always zero during the whole time-optimal 
evolution. 

On the other hand, when there is some initial entanglement between qubits 1 and 3, then we have a 
non trivial time-optimal evolution of the $\tau_{13}(\tau)$ and $\tau_{123}(\tau)$ tangles.

B2) 
When the initial state belongs to the class $B2$ in eq.~\eqref{eq-B2} 
of bi-separable states, from eqs. \Ref{uopt}, \Ref{13tangle}, 
\Ref{123tangle}, \Ref{psitau}, \Ref{hypdet}, \Ref{detrho123} and \Ref{abisep2} and we get the following time-optimal evolutions:
\begin{align}
\tau_{13_{B2}}(\tau)&=|a_{2}^*a_{3}+\hat B_0^2s_{2}s_{3}|^2,
\label{13tanbisep2}
\\
\tau_{123_{B2}}(\tau)&=\frac{\hat B_0^2}{4}|a_{2}s_{3}-a_{3}s_{2}|^2=1-\tau_{13_{B2}}(\tau).
\label{123tanbisep2}
\end{align}
 
W) 
The case of an initial $W$ state is similar and we find that $\tau_{13_{W}}(\tau)=(4/9)\tau_{13_{B2}}(\tau)$ and 
$\tau_{123_{W}}(\tau)=(4/9)\tau_{123_{B2}}(\tau)$.

GHZ) 
Finally, for the class of fragile, fully entangled $GHZ$ states, the
time-optimal evolution of the tangles is found to be: 
\begin{align}
\tau_{13_{GHZ}}(\tau)&=\hat B_0^2|a_{1}s_{4}-a_{4}s_{1}|^2,
\label{13tanghz}
\\
\tau_{123_{GHZ}}(\tau)&=\frac{1}{4}|a_{1}^*a_{4}+\hat B_0^2s_{1}s_{4}|^2=1-\tau_{13_{GHZ}}(\tau).
\label{123tanghz}
\end{align}

\section{Time-optimal evolution of Entanglement}

In this section, we shall study in more detail 
the behavior of the time-optimal evolution of the tangles $\tau_{13}$
and 
$\tau_{123}$ which witness the entanglement between  
qubits 1 and 3. 

We use of the
formulas \Ref{si}, \Ref{ai} and \Ref{bi} explicitly in equations 
\Ref{13tanbisep2}-\Ref{123tanghz}. 
As the tangles are always related by $\tau_{13}(\tau)=1-\tau_{123}(\tau)$, we limit ourselves to the study of the $\tau_{13}$
tangle.

B2-W) For the case of an initial bi-separable state of class $B2$ (and similarly for initial states of the class $W$), from 
\Ref{ai} and \Ref{13tanbisep2} we obtain:
\begin{align}
\tau_{13_{B2}}(\tau)&=[c_2c_3+\hat\omega_K^2(\cos^2\phi +b_2b_3)s_2s_3]^2
\nonumber \\
&+\hat\omega_K^2(b_2s_2c_3-b_3s_3c_2)^2.
\label{tanb2}
\end{align}
We now proceed by further optimizing the tangle $\tau_{13_{B2}}(\tau)$ as a function of the time $\tau$ and of the unknown 
constants of the
motion for the time-optimal quantum trajectory, i.e. $\hat B_0, \hat B_z, \hat \Omega$, for a given ratio of the couplings 
$K$ and for fixed energy $\hat \omega$.
In particular, we look for the time $\tau_\ast$ at which the tangle first reaches its maximum value 
$\tau_{13_{B2}}|_{\mathrm{max}}=1$ by imposing
that the partial derivatives of the tangle with respect to $\tau$ and the other constants of the motion (expressed in terms of
$\phi$ and $\hat \Omega$) vanish and that the determinant of the Hessian matrix is negative.
After a long and tedious but simple algebra we find that ``optimal''
time at which $\tau_{13_{B2}}(\tau)$ first reaches its maximum  
depends on the value of the available energy $\hat \omega^2$ and on the ratio $K$ between the couplings in the
Ising Hamiltonian. 
In more details, we have that:  
\begin{align}
\tau_{\ast_{B2}}=\frac{\sqrt{3}}{4}\frac{\pi}{|1-K|},
\label{tanb2opt1}
\end{align}
when $1<\hat\omega^2< 29/16$ and $|K|<K_{1+}$, or when $\hat\omega^2> 29/16$ and
$K_{1-}< K < K_{2-}$ or $K_{2+} < K < K_{1+}$, where we have defined $K_{1\pm}:= \pm \sqrt{\hat\omega^2-1}$ 
and $K_{2\pm}:= (1/4)[13/4 ~\pm \sqrt{3(\hat\omega^2-29/16)}]$. 
The duration \Ref{tanb2opt1} of the optimal quantum evolution for the bi-partite entanglement is minimal for 
a ratio of the Ising 
couplings $K\rightarrow \mp K_{1-}$, and maximal for a ratio of the Ising couplings $K\rightarrow K_{1+}$. 
Furthermore, in this case we obtain the optimal magnitudes and frequency for the magnetic field as:
\begin{align}
|\hat B_{0_{B2}}|&=\frac{2}{\sqrt{3}}|K-1|,
\\
|\hat B_{z_{B2}}|&=\sqrt{\hat\omega^2-\frac{7}{3}K^2+\frac{8}{3}K-\frac{49}{21}},
\\
\hat\Omega_{B2}&=2\left [K-1\pm \sqrt{\hat\omega^2-\frac{7}{3}K^2+\frac{8}{3}K-\frac{49}{21}}\right ].
\label{bzb0omoptb21}
\end{align}

Instead, we have that:
\begin{align}
\tau_{\ast_{B2}}=\frac{\pi}{\sqrt{\hat\omega^2-2K}},
\label{tanb2opt2}
\end{align}
when $\hat\omega^2> 29/16$ and $K_{2-}< K < K_{2+}$.
Within this range of allowed energies and $K$, the duration \Ref{tanb2opt2} of the optimal quantum evolution for the bi-partite 
entanglement is minimal for a ratio of the Ising couplings $K\rightarrow K_{2-}$, and maximal for a ratio of the Ising couplings $K\rightarrow K_{2+}$. 
In this case, the optimal values for the magnetic field are:
\begin{align}
|\hat B_{0_{B2}}|&=|\hat\omega_K|,
\\
|\hat B_{z_{B2}}|&=|\hat\Omega|=0.
\label{bzb0omoptb22}
\end{align} 
In Figs. 1-2 we plot the 2-tangle and the 3-tangle as a function of time when the quantum system is in the initial 
bi-separable state $B2$ and for the value of energy $\hat\omega=\sqrt{6}$.
For this value of energy, we can compute $K_{1\pm}=\pm 2.24$ and $K_{2+}\simeq 1.70$, $K_{2-}\simeq -0.007$.
To exemplify the values of the couplings, we consider two models proposed in \cite{nimbalkar} for nuclear magnetic resonance (NMR) experiments, i.e.: i) the $H-N-H$ chain in the molecule of ethanamide, for which 
$J_{12}=J_{23} \simeq 88.05$ Hz and therefore $K=1$; ii) the $P-F-H$ chain in the molecule of diethylfluoromethylphosphonate, for which $J_{12}\simeq 46$ Hz, $J_{23}\simeq 73.1$ Hz and therefore $K=1.59$.
For both $K=1$ and $K=1.59$ we are in the situation depicted by formula \Ref{tanb2opt2}, and  
the law of quantum evolution of the tangle is explicitly given by 
$\tau_{13_{B2}}(\tau)=1-4[(1-K)^2\hat\omega_K^2/(\hat\omega^2-2)^2]\sin^4[\sqrt{\hat\omega^2-2K}\tau]$.
We notice that the case of equal Ising couplings, i.e. case i) with $K=1$, residual entanglement between the indirectly coupled qubits is constant and always maximal, equal to one, while the tripartite entanglement is always zero.
For the case ii) of unequal couplings, instead, the bipartite entanglement has a periodic behavior, starting from the 
maximum equal to one at $\tau=0$, reaching a minimum of approximately $0.57$ at $\tau=\tau_{\ast_{B2}}/2$, 
rising again to the maximum of one at $\tau=\tau_{\ast_{B2}}$ and so on.

GHZ) Let us now turn to the case of an initial state belonging to the $GHZ$ class.
From \Ref{ai} and \Ref{13tanghz} we obtain:
\begin{align}
\tau_{13_{GHZ}}(\tau)&=\hat \omega_K^2\cos^2\phi [(c_1s_4-c_4s_1)^2
\nonumber\\
&+\hat\omega^2_K(b_4-b_1)^2(s_1s_4)^2].
\label{tanb2}
\end{align}
In this case, the ``optimal'' time at which $\tau_{13_{GHZ}}(\tau)$ first reaches its maximum is found to be:
\begin{align}
\tau_{\ast_{GHZ}}=\frac{\sqrt{2}}{4}\frac{\pi}{|1+K|},
\label{tanghzopt1}
\end{align}
provided that $\hat\omega^2 >  3/2$ and $K_{-} < K <  K_{+}$,
where we have defined $K_{\pm}:= (1/2)[-1\pm \sqrt{2\hat\omega^2-3}]$.
For the $GHZ$ initial states, the duration \Ref{tanghzopt1} of the optimal quantum evolution for the bi-partite 
entanglement is minimal for a ratio of the Ising couplings $K\rightarrow K_{+}$, and maximal for a ratio of the Ising couplings $K\rightarrow K_{-}$ (and it diverges if $\hat\omega^2=2$, and $K\rightarrow K_- =-1$). 
The associated optimal magnitudes and frequency for the magnetic field in the $GHZ$ case are:
\begin{align}
|\hat B_{0_{GHZ}}|&=|1+K|,
\\
|\hat B_{z_{GHZ}}|&=\frac{|\hat\Omega_{GHZ}|}{2}=\sqrt{\hat\omega^2-2(K^2+K+1)}.
\label{bzb0omoptghz}
\end{align}

In Figs. 3-4 we plot the 2-tangle and the 3-tangle as a function of time when the quantum system is in the initial 
$GHZ$ state and for the value of energy $\hat\omega=\sqrt{14}$.
For this energy we have that $K_{-}=-3$ and $K_{+}=2$.
Therefore, we can still use the NMR models i) and ii) of Figs. 1-2, with $K=1$ and $K=1.59$, respectively.
Here the explicit analytical formula for the time-optimal evolution of the entanglement is given by
$\tau_{13_{GHZ}}(\tau)=\sin^4[\sqrt{2}(1+K)\tau]$.
Now the periodical behavior is present for both values of $K$, and the system oscillates between the initial maximal
bipartite entanglement and zero entanglement.

\section{Discussion}

We have analytically investigated the problem of the time-optimal 
unitary evolution of (tripartite) entanglement, a fundamental resource in
quantum computation and quantum information \cite{horodecki}. 
How to robustly create entanglement in the shortest possible way (to fight decoherence etc...)
is a crucial task which is the subject of several efforts in the literature (see Introduction).
Our method of analysis based on the quantum brachistochrone (QB), developed by the
present authors, is very general and can be applied to arbitrary quantum systems. 
The model that we considered consists of indirectly coupled qubits via an intermediate qubit that
is directly controllable. An example of a concrete physical system which realizes the model 
is a trilinear Ising chain with unequal interaction couplings, with the middle spin controlled by a local magnetic field. 
The entanglement is quantified by the 2-tangle between the two qubits at the end of the chain.
Using the formalism of the QB with the constraint of a fixed energy available, 
we have analytically found the time-optimal unitary evolution law for the Ising Hamiltonian plus
the local control and we substituted it in the formula for the 2-tangle.
The initial boundary condition for the QB is chosen among all possible sets of tripartite quantum pure states,
i.e. fully separable, bi-separable, $W$ and $GHZ$ states.
The integrals of the motion in the QB are determined imposing that the 2-tangle reaches its maximum in the shortest time possible, which we call $\tau_\ast$.
Entanglement is found to have a non trivial role during the time-optimal unitary evolutions of $W$ and $GHZ$ initial quantum states, and of the bi-separable initial state in which the indirectly coupled qubits have a nonzero value of the 2-tangle.
The optimal time $\tau_\ast$  also sets the time-scale for the duration of the significant role of the entanglement,
and it is a function of the ratio $K$ between the interaction couplings in the Ising Hamiltonian and of the energy
available in the experiment.
The monogamy of entanglement shows neatly in the anti-correlation of the tripartite entanglement, quantified by the 
3-tangle, with the bipartite entanglement, quantified by the 2-tangle.
The QB method has been used to study a physical example which is a typical and interesting scenario in quantum information processing 
and which had been investigated under different perspectives in the previous literature \cite{khaneja}-\cite{vandamme}.
We extended this investigation to the important case of entanglement, and our work is a rare example where this kind of analysis is done in a completely analytical manner.
It is well known in the theory of quantum optimal control that going beyond and considering the analytical description of more complex (e.g., with more than 2-3 qubit) systems is 
an extremely challenging task (though several results on higher dimensional models exist, they are all based on numerical approaches). 
Nevertheless, it is our intention to try and extend the analysis presented here at least to the case when, e.g.,  coherent control is possible on all
the qubits in the chain, to longer chains of qubits (with certain symmetries), and to study the QB evolution of the truly non classical 
correlations via their proper measure, quantum discord \cite{ollivier}.

\vspace{1cm}

\section*{ACKNOWLEDGEMENTS}
A.C. acknowledges the support from the MIUR of Italy under the
program ``Rientro dei Cervelli''.
T.K acknowledges the support from 
MEXT-Supported Program for the Strategic Research Foundation at
Private Universities ``Topological Science'' and from 
Keio University G-SEC Creativity Initiative ``Quantum Community''. 


\section{Appendix}

For a tripartite system 123 in a pure state
$\ket{\psi}=\sum_{i=0}^7a_i\ket{i}$ expanded in the basis
$\{\ket{0}:=\ket{000}$, $\ket{1}:=\ket{001}$, $\ket{2}:=\ket{010}$,
$\ket{3}:=\ket{011}$, $\ket{4}:=\ket{100}$, $\ket{5}:=\ket{101}$,
$\ket{6}:=\ket{110}$,  
$\ket{7}:=\ket{111}\}$, 
Cailey's hyperdeterminant for the matrix of the coefficients $a_i$s is defined as:
\begin{align}
\mathrm{HypDet}(a)&:=[(a_0a_7)^2 +(a_1a_6)^2 + (a_2a_5)^2 + (a_3a_4)^2]
\nonumber \\
& -2[(a_0a_7+a_1a_6)(a_2a_5+a_3a_4)
\nonumber \\
&+a_0a_1a_6a_7 +a_2a_3a_4a_5]
\nonumber \\
&+4(a_0a_3a_5a_6+a_1a_2a_4a_7).
\label{hypdet}
\end{align}
More explicitly, since:
\begin{align}
\mathrm{Det}(\rho_1)&=|a_0|^2(|a_5|^2+|a_6|^2+|a_7|^2) 
\nonumber \\
&+|a_1|^2(|a_4|^2+|a_6|^2+|a_7|^2)
\nonumber \\
&+|a_2|^2(|a_4|^2+|a_5|^2+|a_7|^2)
\nonumber \\
&+|a_3|^2(|a_4|^2+|a_5|^2+|a_6|^2)
\nonumber \\
&-2\mathrm{Re}(a_0a_5a_1^*a_4^*+a_0a_6a_2^*a_4^*
\nonumber \\
&+a_0a_7a_3^*a_4^*+a_1a_6a_2^*a_5^*
\nonumber \\
&+a_1a_7a_3^*a_5^*+a_2a_7a_3^*a_6^*),
\label{rho1}
\end{align}
\begin{align}
\mathrm{Det}(\rho_2)&=|a_0|^2(|a_3|^2+|a_6|^2+|a_7|^2) 
\nonumber \\
&+|a_1|^2(|a_2|^2+|a_6|^2+|a_7|^2)
\nonumber \\
&+|a_4|^2(|a_4|^2+|a_3|^2+|a_7|^2)
\nonumber \\
&+|a_5|^2(|a_4|^2+|a_3|^2+|a_6|^2)
\nonumber \\
&-2\mathrm{Re}(a_0a_3a_1^*a_2^*+a_0a_6a_2^*a_4^*
\nonumber \\
&+a_0a_7a_2^*a_5^*+a_1a_6a_3^*a_4^*
\nonumber \\
&+a_1a_7a_3^*a_5^*+a_4a_7a_5^*a_6^*),
\label{rho2}
\end{align}
and
\begin{align}
\mathrm{Det}(\rho_3)&=|a_0|^2(|a_3|^2+|a_5|^2+|a_7|^2) 
\nonumber \\
&+|a_2|^2(|a_1|^2+|a_5|^2+|a_7|^2)
\nonumber \\
&+|a_4|^2(|a_1|^2+|a_3|^2+|a_7|^2)
\nonumber \\
&+|a_6|^2(|a_1|^2+|a_3|^2+|a_5|^2)
\nonumber \\
&-2\mathrm{Re}(a_0a_3a_1^*a_2^*+a_0a_5a_1^*a_4^*
\nonumber \\
&+a_0a_7a_1^*a_6^*+a_2a_5a_3^*a_4^*
\nonumber \\
&+a_2a_7a_3^*a_6^*+a_4a_7a_5^*a_6^*),
\label{rho3}
\end{align}
we obtain:
\begin{align}
\mathrm{Det}(\rho_1)-\mathrm{Det}(\rho_2)+&\mathrm{Det}(\rho_3)=2(|a_0|^2|a_5|^2+|a_1|^2|a_4|^2
\nonumber \\
&+|a_2|^2|a_7|^2+|a_3|^2|a_6|^2)
\nonumber \\
&-2\mathrm{Re}[a_0a_7a_1^*a_6^*+a_2a_5a_3^*a_4^*
\nonumber \\
&-(a_0a_7-a_1a_6)(a_2^*a_5^*-a_3^*a_4^*)]
\nonumber \\
&-4\mathrm{Re}(a_0a_5a_1^*a_4^*+a_2a_7a_3^*a_6^*).
\label{detrho123}
\end{align}

Given the time-optimal unitary evolution \Ref{psitau}, we can compute the following nonzero time-dependent amplitudes
(all modulo $\exp[-i\Omega\tau/2]$):

S) representative of the class of fully separable initial states, $\ket{\psi_S}$:
\begin{align}
(a_S)_ 0(\tau)&=a_{1}^*,
\nonumber \\
(a_S)_2(\tau)&=-i\hat B_0e^{i\theta}s_{1};
\label{asep}
\end{align}

B1) representative of the class of bi-separable initial states, $\ket{\psi_{B1}}$:
\begin{align}
(a_{B1})_0(\tau)&=-i\frac{\hat B_0}{\sqrt{2}}e^{-i\theta_0}s_{1},
\nonumber \\
(a_{B1})_1(\tau)&=\frac{a_{2}^*}{\sqrt{2}},
\nonumber \\
(a_{B1})_2(\tau)&=\frac{e^{i\hat \Omega\tau}}{\sqrt{2}}a_{1},
\nonumber \\
(a_{B1})_3(\tau)&=-i\frac{\hat B_0}{\sqrt{2}}e^{i\theta}s_{2};
\label{abisep1}
\end{align}

B2) representative of the class of bi-separable initial states, $\ket{\psi_{B2}}$: 
\begin{align}
(a_{B2})_1(\tau)&=\frac{a_{2}^*}{\sqrt{2}},
\nonumber \\
(a_{B2})_3(\tau)&=-i\frac{\hat B_0}{\sqrt{2}}e^{i\theta}s_{2},
\nonumber \\
(a_{B2})_4(\tau)&=\frac{a_{3}^*}{\sqrt{2}},
\nonumber \\
(a_{B2})_6(\tau)&=-i\frac{\hat B_0}{\sqrt{2}}e^{i\theta}s_{3};
\label{abisep2}
\end{align}

B3) representative of the class of bi-separable initial states, $\ket{\psi_{B3}}$: 
\begin{align}
(a_{B3})_0(\tau)&=-i\frac{\hat B_0}{\sqrt{2}}e^{-i\theta_0}s_{1},
\nonumber \\
(a_{B3})_2(\tau)&=\frac{e^{i\hat \Omega\tau}}{\sqrt{2}}a_{1},
\nonumber \\
(a_{B3})_4(\tau)&=\frac{a_{3}^*}{\sqrt{2}},
\nonumber \\
(a_{B3})_6(\tau)&=-i\frac{\hat B_0}{\sqrt{2}}e^{i\theta}s_{3};
\label{abisep3}
\end{align}

W) representative of the class of  $W$ initial states, $\ket{\psi_W}$:
\begin{align}
(a_W)_0(\tau)&=-i\frac{\hat B_0}{\sqrt{3}}e^{-i\theta_0}s_{1},
\nonumber \\
(a_W)_1(\tau)&=\frac{a_{2}^*}{\sqrt{3}},
\nonumber \\
(a_W)_2(\tau)&=\frac{e^{i\hat \Omega\tau}}{\sqrt{3}}a_{1},
\nonumber \\
(a_W)_3(\tau)&=-i\frac{\hat B_0}{\sqrt{3}}e^{i\theta}s_{2},
\nonumber \\
(a_W)_4(\tau)&=\frac{a_{3}^*}{\sqrt{3}},
\nonumber \\
(a_W)_6(\tau)&=-i\frac{\hat B_0}{\sqrt{3}}e^{i\theta}s_{3};
\label{aw}
\end{align}

GHZ) representative of the class of  GHZ initial states, $\ket{\psi_{GHZ}}$:
\begin{align}
(a_{GHZ})_0(\tau)&=\frac{a_{1}^*}{\sqrt{2}},
\nonumber \\
(a_{GHZ})_2(\tau)&=-i\frac{\hat B_0}{\sqrt{2}}e^{i\theta}s_{1},
\nonumber \\
(a_{GHZ})_5(\tau)&=-i\frac{\hat B_0}{\sqrt{2}}e^{-i\theta_0}s_{4},
\nonumber \\
(a_{GHZ})_7(\tau)&=\frac{e^{i\hat \Omega\tau}}{\sqrt{2}}a_{4}.
\label{aghz}
\end{align}

\bibliographystyle{alpha}

\begin{figure}[H]
 \begin{center}
  \resizebox{8cm}{!}{\includegraphics{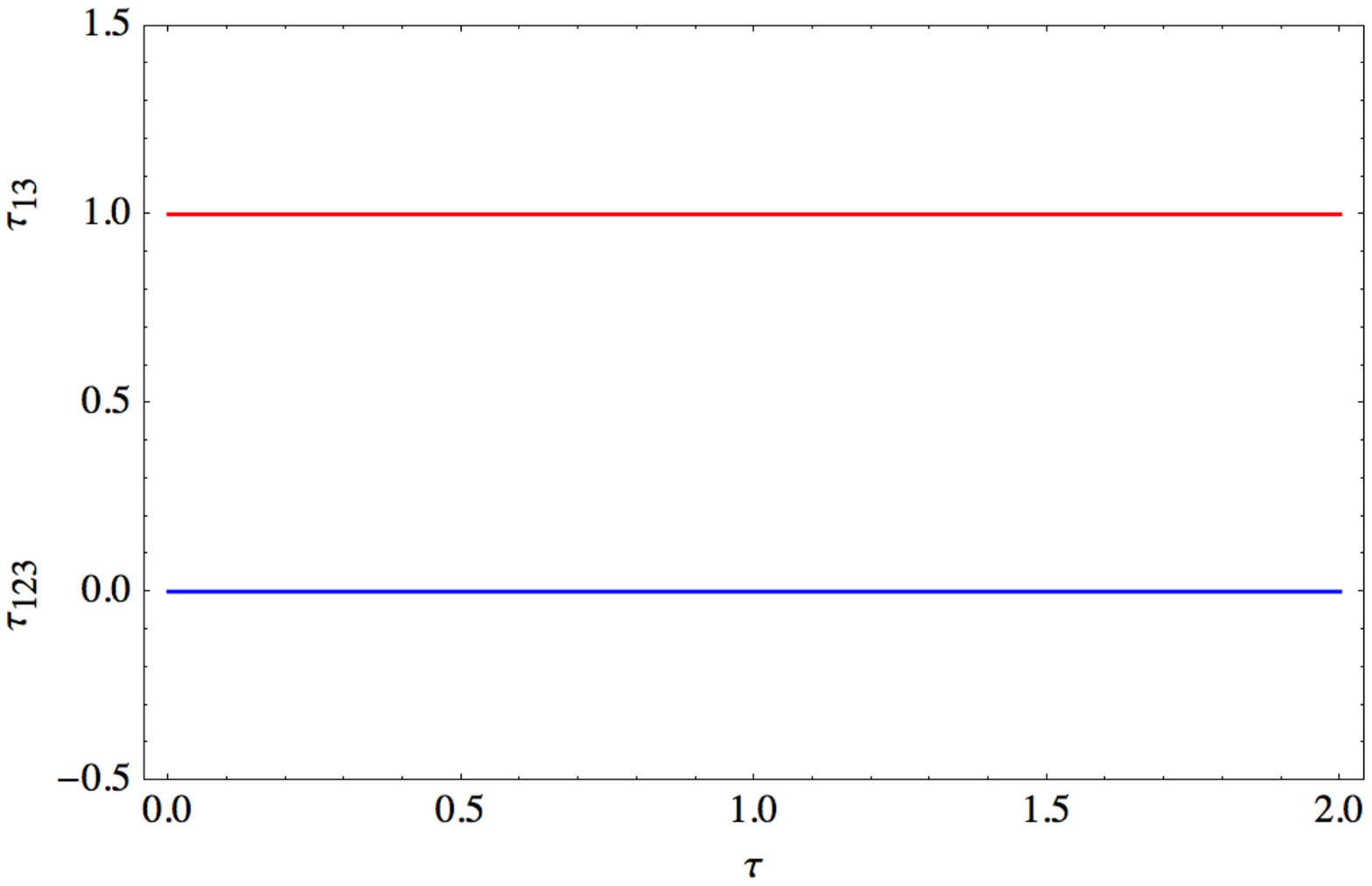}}
  \caption{The 2-tangle (red curve) and the 3-tangle (blue curve) as a function of time 
  for the $B2$ initial states and the coupling ratio $K=1$.}
 \end{center}
\end{figure}
\begin{figure}[H]
 \begin{center}
  \resizebox{8cm}{!}{\includegraphics{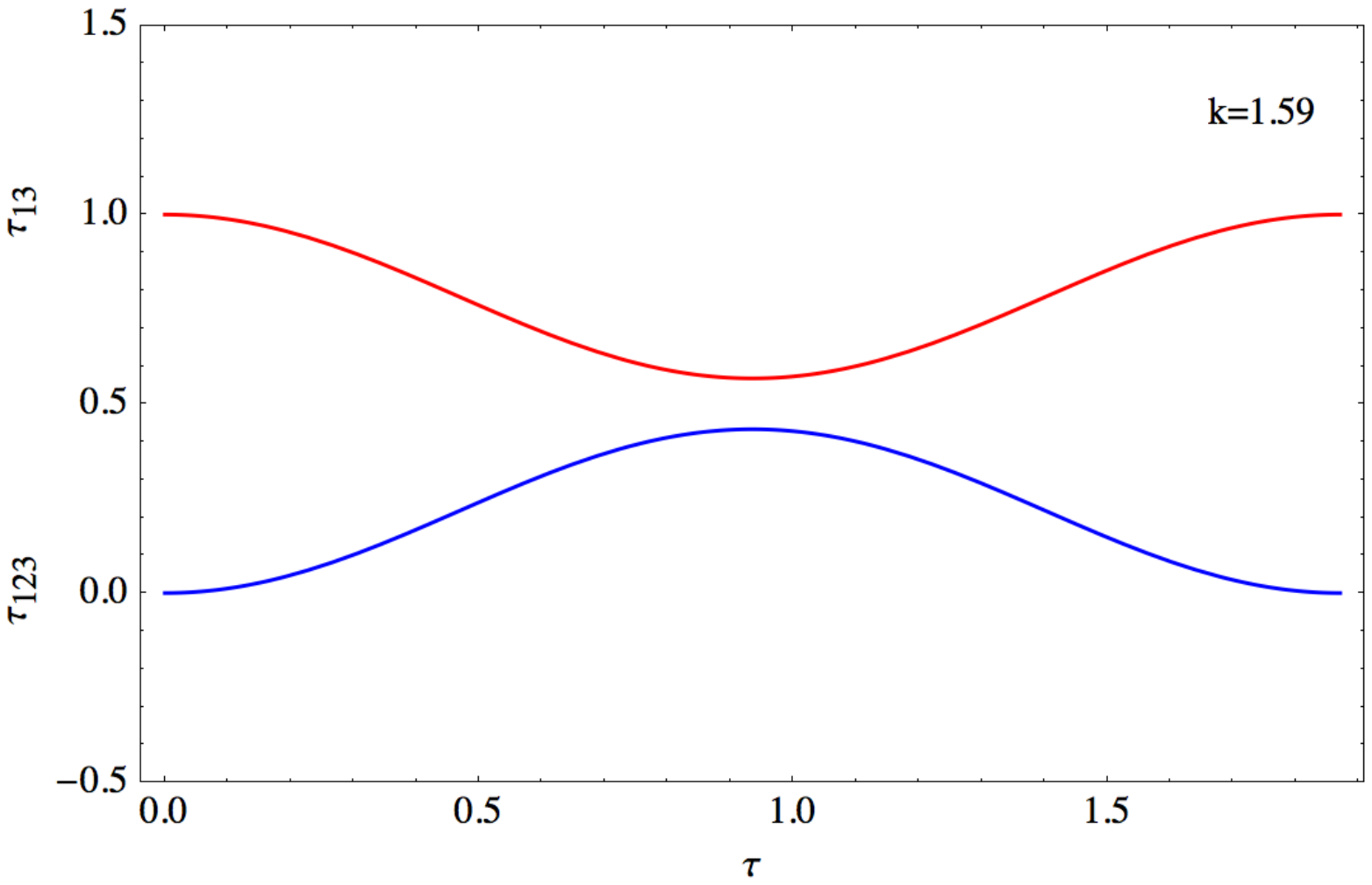}}
  \caption{The 2-tangle (red curve) and the 3-tangle (blue curve) as a function of time  for the $B2$ initial states and the coupling ratio $K=1.59$. }
 \end{center}
\end{figure}
\begin{figure}[H]
 \begin{center}
  \resizebox{8cm}{!}{\includegraphics{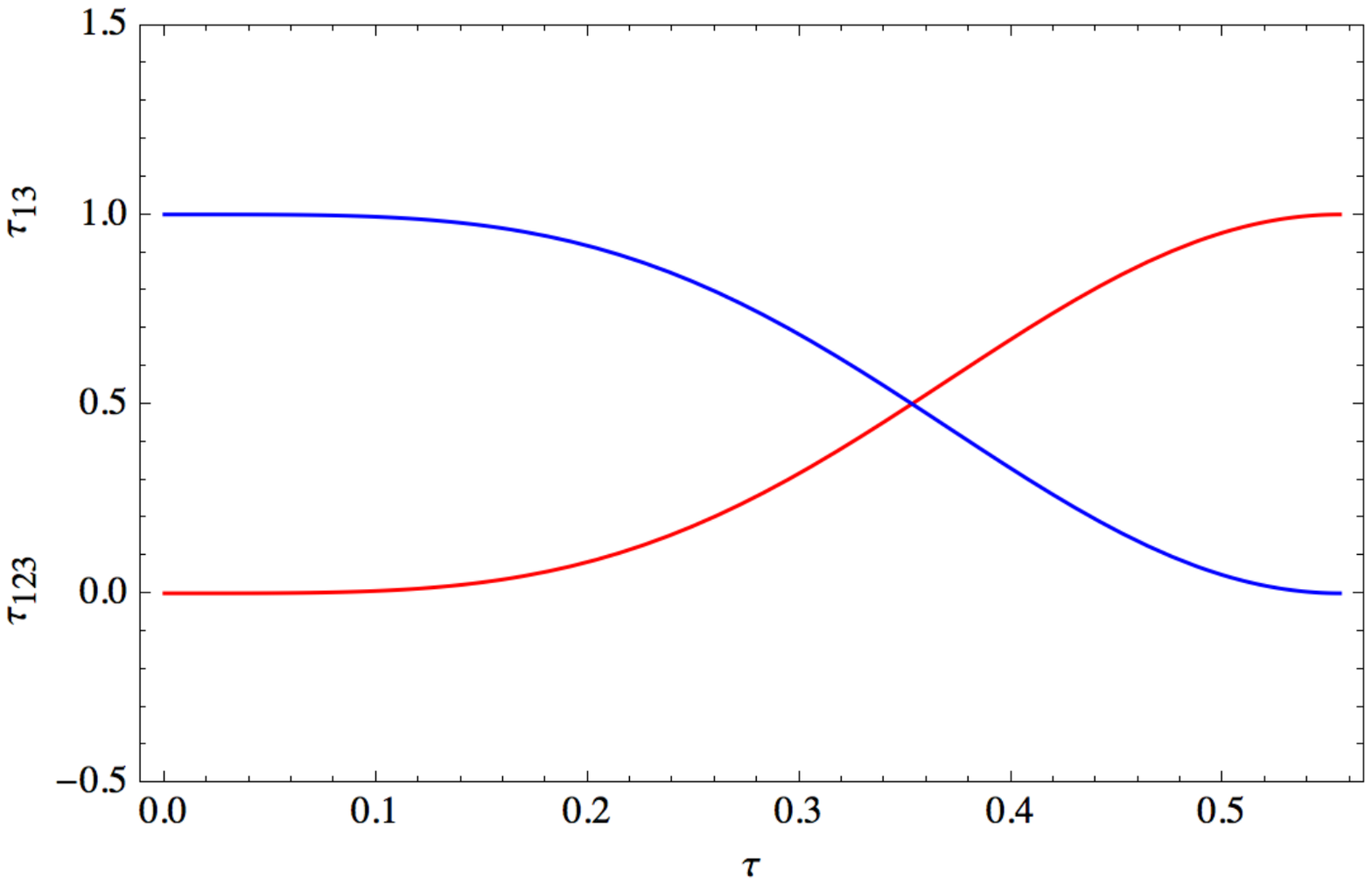}}
  \caption{The 2-tangle (red curve) and the 3-tangle (blue curve) as a function of time for the $GHZ$ initial states and the coupling ratio $K=1$.}
 \end{center}
\end{figure}
\begin{figure}[H]
 \begin{center}
  \resizebox{8cm}{!}{\includegraphics{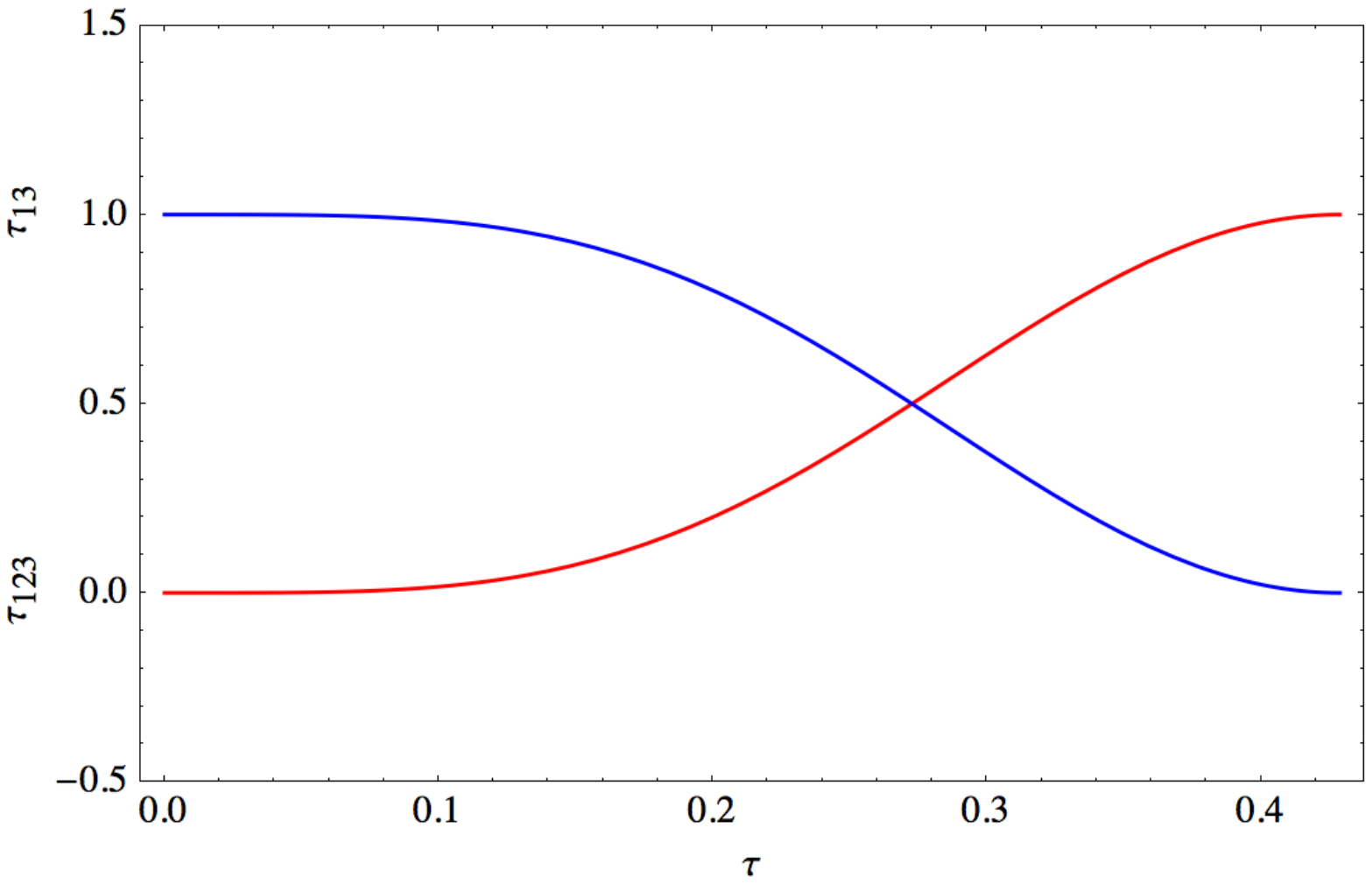}}
  \caption{The 2-tangle (red curve) and the 3-tangle (blue curve) as a function of time for the $GHZ$ initial states and the coupling ratio $K=1.59$. }
 \end{center}
\end{figure}

\end{document}